\documentclass[journal]{IEEEtran}
\usepackage{graphicx}
\usepackage{subcaption}
\usepackage{multirow} 
\usepackage{siunitx} 
\usepackage{helvet}
\usepackage{titlesec}

\titleformat{\chapter}
  {\normalfont\fontsize{16}{19}\sffamily\bfseries}
  {\thechapter}
  {1em}
  {}

\titleformat{\section}
  {\normalfont\fontsize{12}{17}\sffamily\bfseries}
  {\thesection}
  {1em}
  {}

\titleformat{\subsection}
  {\normalfont\fontsize{12}{17}\sffamily\bfseries\slshape}
  {\thesubsection}
  {1em}
  {}
\renewcommand{\thesection}{\arabic{section}}

\usepackage{cite}

\ifCLASSINFOpdf
\else
\fi
\usepackage{fixltx2e}

\usepackage{stfloats}
\usepackage{authblk}

\begin{document}
%

\title{Psychologically-Inspired Music Recommendation System}
%
%
%

\author{Danila~Rozhevskii,
        Jie~Zhu,
        Boyuan~Zhao
\thanks{J. Wang is with Penn State University}}
\affil{Department of Computer Science, Penn State University}

\markboth{DS340W Term Paper}%
{Shell \MakeLowercase{\textit{et al.}}: Bare Demo of IEEEtran.cls for IEEE Journals}

\maketitle

\begin{abstract}
In the last few years, automated recommendations systems have been a major focus in the music field, where companies such as Spotify, Amazon, and Apple are competing in the ability to generate the most personalized music suggestions for their users. One of the challenges developers still fail to tackle is taking into account the psychological and emotional aspects of the music. Our goal is to find a way to integrate users' personal traits and their current emotional state into a single music recommendation system with both collaborative and content-based filtering. We seek to relate the personality and the current emotional state of the listener to the audio features in order to build an emotion-and-personality aware MRS. We compare the results both quantitatively and qualitatively to the output of the traditional MRS based on the Spotify API data to understand if our advancements make a significant impact on the quality of music recommendations.
\end{abstract}


\IEEEpeerreviewmaketitle

\section{Introduction}
Since the development of audio and video compression techniques and transition to mobile platforms, both music and movie industries adopted to sell and consume their products digitally. That’s why the typical concept of owning music switched from buying a physical collection to paying a streaming subscription~\cite{deldjoo2021content}. While systems in the movies domain have been already thoroughly researched with common use cases such as typical user-item recommendations, the music systems require more in-depth consideration of various more diverse scenarios. Having huge and diverse libraries of songs of all kinds of genres, the personal recommendation algorithm is tasked not with “selling” one particular song, but with providing an almost infinite loop of recommendations~\cite{schedl2017new}.

In 2022, the number of research done on MRS (Music Recommendation Systems) has been increasing drastically with people coming up with new ways to increase the quality of song recommendations, which started to introduce both major opportunities and challenges for recommendation systems in general. There are some features of music recommendations that were highlighted in the ACM Recommender Systems 2017 conference that distinguishes them from other digital items, such as movies, books, or products~\cite{abel2017recsys}. 

First of all the duration of a song is a few times shorter than that of a movie, about 3 to 5 min long, which makes it easily disposable. Therefore, the second feature is the large variety magnitude of items, which essentially means that the size of music catalogs far exceeds those of movies or books. Recommendation algorithms have to work tens of millions of music pieces as compared to tens of thousands of movies. Finally, there are unique trends in users music consumption behavior. For example, by listening music in the background, some people might not consider skipping repeating tracks, which can give a wrong positive signal to the algorithm. In our paper, we focus on the emotional aspect of music recommendations, which is usually neglected in the current MRS and could potentially allow system to be emotion-aware. If we manage to achieve that, we could bring music recommendation systems to a whole new state-of-the-art level~\cite{schedl2018current}.

The field of personalized music recommendations still lacks attention to the psychological and emotional aspects of music. In some situations, it can be crucial to consider these intrinsic characteristics of a listener to provide him or her with the best listening experience. During one of the studies, it was found that music preferences were one of the strongest factors to determining one’s personality. Researchers claimed that music preferences  can reveal more information about personality traits than other lifestyle and leisure domains, with the exception of hobbies~\cite{rentfrow2003re}. Current approaches experience difficulty with combining emotional features of the music to the listener's personality due to the fact that people’s perception of music genres is different and can be misinterpreted by the system. In our work, we try to find a way to create a more genuine and instant connection between a listener and recommended music in a particular moment of time. 

Our project is aimed to further explore the analysis of audio features of songs combined with their emotional message and user current emotional state to develop a psychologically inspired music recommendation. This can be achieved by the combination of existing approaches with the focus on connecting listener emotional state with music emotion tags. The collaborative filtering as well as popularity methods help us consider the songs with the highest user rating that naturally puts them on the spot. The content-based methods allow us to analyze the audio features of the songs such as their frequencies to classify them into groups of similar sounding. The information on the user is collected through a short survey and analysis of the user’s playlist to build the current physiological and emotional state.

\section{Music Recommendations}
Music is a language; it is an expression of emotions and is more subtle and powerful than words. Music is also a bridge to emotions, appealing directly to the emotional and social processing centers of the human brain through the auditory channel, causing each of us to empathize with each other.
Ball University researchers also believe that music is an important channel for evoking human emotions.
According to Damasio~\cite{damasio1999feeling}, Music has different effects on humans in three dimensions: subjective feeling, behavioral changes, and physiological responses. 
Subjective feeling refers to the emotional changes that people experience when they hear music. For example, when people are in trouble in life, they are usually very depressed, and if they hear some slow songs, their emotions will be driven and they will think that the song matches their emotions at the moment, thus achieving an emotional resonance. behavioral changes are actually very common in life.

For example, when we go to a live house to bounce, the DJ will usually choose some fast-paced songs to mobilize the atmosphere, and when the audience hears these songs with a strong sense of rhythm, their bodies will unconsciously dance along. physiological responses refer to the stimulation of the human nervous system by music, resulting in sweating and heartbeat changes. physiological responses refer to the stimulation of the human nervous system that results in sweating and changes in heart rate. This shows that music can have different effects on listeners at both physical and psychological levels. In short, listeners' choice of music is based on two major aspects: emotional parameters and lifestyle." Numerous associations existed between musical preference and these aspects of participants' lifestyle"~\cite{north2007lifestyle}. 

In this paper, we hope to recommend more suitable songs to listeners through a more accurate algorithm. On the one hand, it can create higher commercial value for music software, and on the other hand, it can maximize the usefulness of different styles, rhythms and genres of music.
In this era of rapid development, every human individual has his or her own habits and style preferences. In the past, each individual has more or less developed his or her own style of music.
In the experiment of Burns et al.~\cite{burns2002effects}, 60 test subjects were asked to listen to songs with different styles. The conclusion indicated that the type of music preferred by the individual gave the testers the best emotional state.

\section{Related Works}
In this section, we are going to cover a couple of projects that have been already implemented with the similar goal of creating an emotion-aware music recommendation system. We aim to analyze them and try to find similar features with our approaches. 

In her recent article et al.~\cite{fanamby2021emotion}, Fanamby Randri talks about her research project that uses behavioral and emotional features of users in order to provide a more dynamic, customized, and tailored experience. The algorithm is based on Deep Reinforcement Learning that uses music emotions and genres as well as information on users such as age, gender, current mood. The data comes from two sources: Dataset on Induced Musical Emotion from Game with a Purpose Emotify and Biometrics for stress monitoring Kaggle dataset. The first dataset of 400 songs was made with a  game where people were matching various songs with the emotion they felt listening to them. The second dataset is publicly available Heart Rate Variability (HRV) and Electrodermal activity (EDA) data collected anonymously~\cite{koldijk2016detecting}. The DRL-based system was built on the PyTorch-based A2C (Advantage Actor-Critic) model, trained on a custom environment developed using the gym library, an OpenAI toolkit specially used in reinforcement algorithm development~\cite{fanamby2021emotion}.The output is measured from the user’s feedback after listening to suggested songs in terms of stress level. Decreased stress level indicates positive feedback, the increase means the opposite.

In another paper, Mikhail Rumiantcev, Oleksiy Khriyenko et el.~\cite{rumiantcev2020emotion}were developing an Emotion Based Music Recommendation system that used a combination of artificial intelligence and music therapy approaches to achieve emotion-driven personalization of their recommendations. First, the system collects user data and builds a general user profile (GUP), then based on that the algorithm computes the initial Music-driven Emotional Model (MEM) of a user. After that, the system implements one of the action modes that involve the transition of the corresponding vector of various music towards the desired point in the user’s emotional space. These modes of this process differ in their intensity by changing the minimum distance between corresponding vectors, so the user’s transition from one emotional state to another could go at different paces. This is exceptionally important in cases where people need to constantly keep a certain level of emotional state such as in late-night driving. In this situation, it would make sense to keep the mode with minimal transition distance, so that new suggested music would be energetically similar and would keep the driver focused for as long as needed.  

\graphicspath{{Pics/}}

\begin{figure*}
  \centering
  \includegraphics[width=1\linewidth]{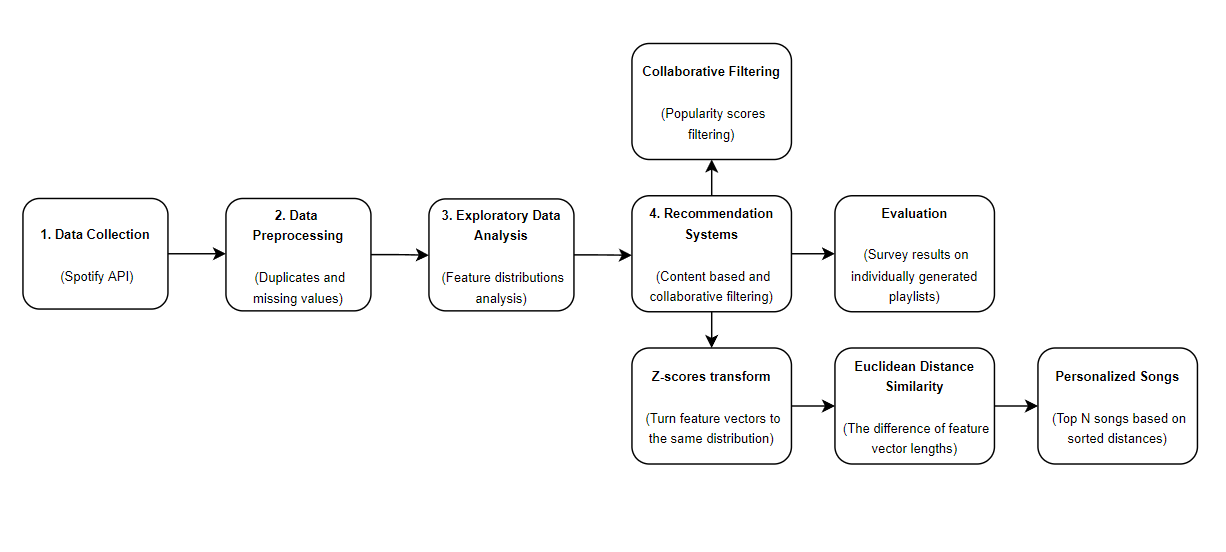}
  \caption{\textbf{Project Diagram}: The pipeline of this project consists of a few general steps: data collection, data pre-processing, exploratory data analysis(EDA), feature engineering, collaborative/content-based filtering (CF), and final results evaluation.}
  \label{figure:sfigure1}
\end{figure*}

\section{Data}
The data used in the project comes from the Spotify API. The reason why we chose to go with Spotify is simply because it is one of the richest and easiest to use music collections in the world with 8 million artists, 1.8 million albums, and over 22 million tracks. Also, it is one of the fastest growing collections, which means that it includes all the latest genres and artists. \par
For computational reasons, we decided to mine over 170,000 songs from over 34,000 different artists from 1921 up to 2020 time period. We thought it would give our algorithm enough variation in data to get high quality recommendation results. Other public data sets that we were able to find did not have enough data to perform good analysis.
We registered a Spotify Developer token and used that to mine the data directly from Spotify API with desired features we chose to work with: valence, energy, tempo, danceability, liveness, and loudness. Spotify does not have filtering tracks by genre, which means that we had to try our best in the algorithm to better differentiate between various genres. 

\section{Psychological Analysis}
Personality is a long-term state of a person's life, but an emotional state is a relatively short-term response compared to personality. People can have many emotional states in a day but maintain the same personality for a long period of time. At the same time, personality can influence the emotional state at different levels, so we try to combine personality and emotional state to more accurately predict the right songs for listeners. \par
In psychology, five personality traits are the five fundamental dimensions of human personality. It contains the most important variables in personality traits. In our project, we will analyze the personality survey reports of different listeners and the usual song lists of these listeners to predict what kind of songs are suitable for different personalities. \par
The five fundamental personality dimensions:
\begin{itemize}
    \item \textbf{Extroversion}:
        Energetic and enthusiastic approach that includes traits such as sociability, assertiveness, confidence, and ambitiousness.
    \item \textbf{Agreeableness}:
        Person's level of altruism, cooperation, willingness to conform to group norms, and warmth of kindness.
    \item \textbf{Consciousness}:
        The ability to control impulses to facilitate goal-directed behaviour. People with this trait are most likely to follow the norms and rules, be highly organized and efficient in planning.
    \item \textbf{Neuroticism}:
        Contrasts emotional stability with feelings of anxiety, nervousness, and depression. People with this trait are self-conscious, moody, impulsive, and prone to stress.
    \item \textbf{Openness}:
        The breadth and depth of one's life, including the originality and complexity of experiences. Individuals high in openness are knowledgeable, perceptive, and analytical; they seek out experiences, and are more artistic and investigative.  
\end{itemize}

\section{Emotional state}
\begin{figure}
  \centering
  \includegraphics[width=.8\linewidth]{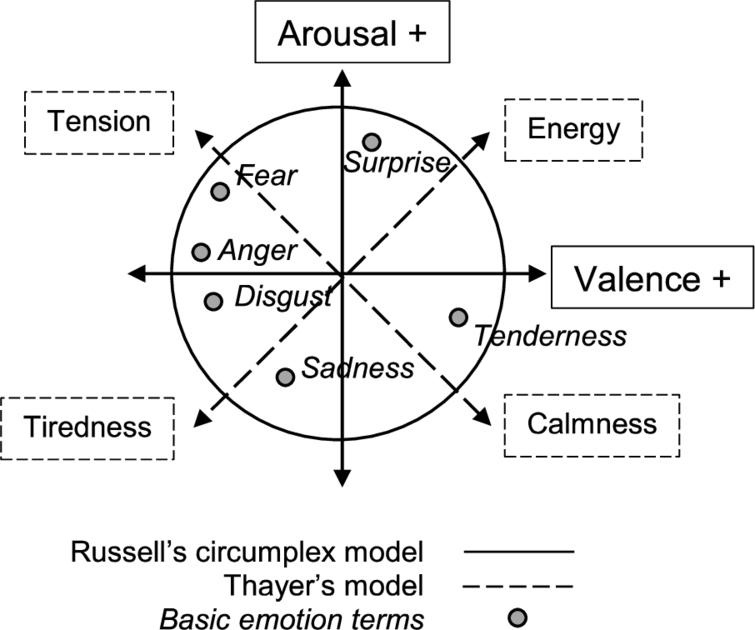}
  \caption{\textbf{Thayer's arousal-valence emotion plane}: Schematic diagram of the two-dimensional models of emotions with common basic emotion categories overlaid (Eerola and Vuoskoski, 2011)}
  \label{fig:sfig2}
\end{figure}

Thayer's arousal-valence emotion plane in Figure 2 divides human emotions into two dimensions: excess-valence. Excite refers to the degree of Arousal of body energy about an emotional state to prepare for action; Valence refers to the degree to which humans like or dislike a person or thing. Its theoretical basis is the separation and activation of positive and negative emotions. \par
After reading previous research, we found Thayers' model was widely applied in Music Research. Many studies have divided complex human emotions into four blocks by Thayer's model, distributed in four quadrants of coordinates. So we can mathematically measure the emotional response people feel when they hear a particular song by referring to Arousal and valence with '+' and '-' For example, '+' and '-' indicate that the audience's emotions are located in the fourth quadrant, which is the state of calmness. Then, when we screen songs' melodies, we can filter out songs with significant frequency peak changes.

\section{Methods}
\subsection{Data pre-processing}
The pipeline of this project, in Figure 1, consists of a few general steps: data pre-processing, feature engineering, collaborative/content-based filtering (CF), and final results evaluation. However, before everything our team had to create a developer account and create an individual project on Spotify For Developers website. That way, we were able to access Spotify API and be able to pull the data.
Data pre-processing mainly consisted of us cleaning up the pulled dataset in order to get rid of duplicate and missing values. A common problem we encountered was that there were a lot of duplicate songs released by same artist, which happened because the track was listed as a single as well as a part of the released album. After all the cleaning, we ended up with over 170,000 unique songs.
 
\subsection{Music features}
In this project, we decided to analyze the following features of songs from Spotify API~\cite{spotify_api}: valence, energy, tempo, danceability, liveness, loudness, and release year. Below, we list feature descriptions as well as our reasoning behind them.
\\~\\
\begin{itemize}
    \item \textbf{Valence}:
        A measure from 0.0 to 1.0 describing the musical positiveness conveyed by a track. Tracks with high valence sound more positive (e.g. happy, cheerful, euphoric), while tracks with low valence sound more negative (e.g. sad, depressed, angry)~\cite{spotify_api}. This is a very important feature, which helps indicate the tracks that are a better fit to the person's mood rather than genre preference. 
    \item \textbf{Energy}:
        Energy is a measure from 0.0 to 1.0 and represents a perceptual measure of intensity and activity. Typically, energetic tracks feel fast, loud, and noisy. For example, death metal has high energy, while a Bach prelude scores low on the scale. Perceptual features contributing to this attribute include dynamic range, perceived loudness, timbre, onset rate, and general entropy~\cite{spotify_api}. Combined with all of our other features, this feature helps us differentiate between old and new songs better. New pop songs and electronic music tend to have high energy values, but can be easily mistaken for older metal songs.
    \item \textbf{Tempo}:
        The overall estimated tempo of a track in beats per minute (BPM). In musical terminology, tempo is the speed or pace of a given piece and derives directly from the average beat duration~\cite{spotify_api}. This feature helps our algorithm to recommend tracks from the similar genres and time periods. For example, Rock music tends to have a BPM range of 110-140, however it can also change depending on a track's release year (see Figure 3).
    \item \textbf{Danceability}:
        Describes how suitable a track is for dancing based on a combination of musical elements including tempo, rhythm stability, beat strength, and overall regularity. A value of 0.0 is least danceable and 1.0 is most danceable~\cite{spotify_api}. 
    \item \textbf{Liveness}:
        Detects the presence of an audience in the recording. Higher liveness values represent an increased probability that the track was performed live. A value above 0.8 provides strong likelihood that the track is live~\cite{spotify_api}. Some of the tracks were performed live, which means they might have different values for the rest of features as well and we have to account for that.
    \item \textbf{Loudness}:
        Almost every artist has his or her own fan base. Some of the more famous artists can have Ins followers in the tens of millions. Many fans will be the first to support an artist's new song when it is released~\cite{spotify_api}. This feature helps a lot with determining the tracks from the past century, since a lot of old songs tend to have lower loudness scores.
\end{itemize}

\begin{figure}
  \centering
  \includegraphics[width=1\linewidth]{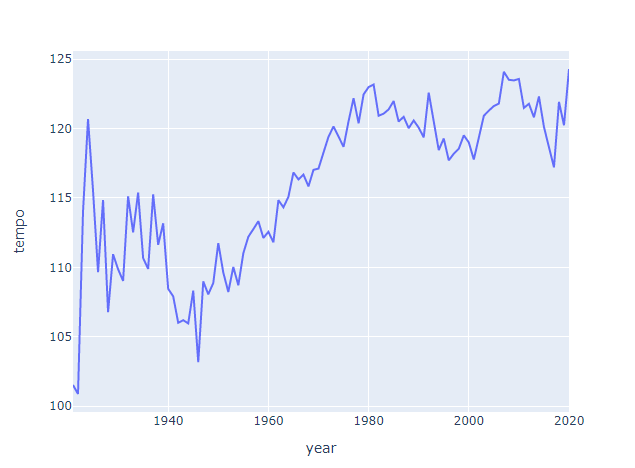}
  \caption{\textbf{Tempo values across years}: the graph is based on the dataset collected for the project. Over the whole time period, the average tempo values of the songs seemed to be increasing.}
  \label{fig:sfig2}
\end{figure}

\subsection{Z-score transformations}

\begin{figure*}
  \centering
  \includegraphics[width=1\linewidth]{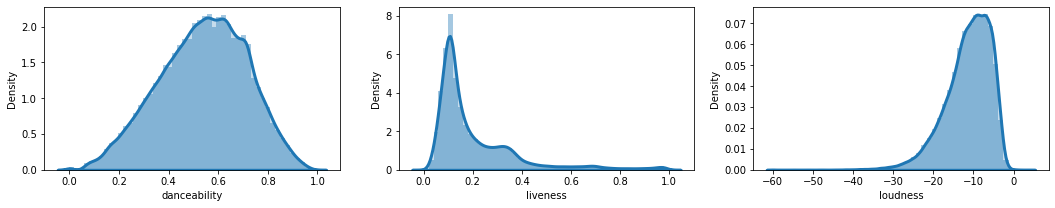}
  \caption{\textbf{Density value distributions}: the density destributions for danceability, liveness, and loudness respectively are all different, which means that we have to apply z-score transformations.}
  \label{fig:sfig2}
\end{figure*}

After performing exploratory data analysis, we discovered that all six features have different density values distributions (see Figure 4), which makes it challenging for us to compare them for different songs. In short, an increase in valence value can have a different impact on our comparison method result than an increase in energy. For that reason, before we compare these features, we have to standardize them to the same normal distribution, which is called a z-score transformation.
\\~\\
Z-score transformation formula:
\begin{center}
$Z=(x-\mu)/\sigma$,
\end{center} 
where $Z$ is the standard value, $x$ is an observed value, $\mu$ is a mean of the feature column, and $\sigma$ is a standard deviation of the feature column.
\\~\\
We apply this formula on the dataset features in order to transform them all to the same value density distribution that we can use for our comparison technique.

\subsection{Euclidean Distance}
After we turned all the features to the common density distribution, we create an accumulative vector variable called 'Mood Vector', which consists of all 6 features as a single one dimensional vector. In order to compare all these features across all the songs in the dataset, we calculated Euclidean distances between a chosen song and all the other songs in our dateset and save them as a 'Distance' variable. 
\\~\\
The Euclidean Distance formula:
\begin{center}
$ d\left(p,q\right) = \sqrt{\sum_{i=1}^{n}\left(q_{i}-p_{i}\right)^2}$,
\end{center} 
where $q_{i},q_{i}$ are two song vectors that have 6 feature values each. The result, $d$, is the length of a line segment between the two vector points.
\\~\\
Finally, we sort the 'Distance' variable by its length and pick the top K songs. These songs are now the result of the content-based filtering performed by our recommendation system.

\subsection{Collaborative filtering}
This technique is widely used in movie recommendations and helps select collection of items that are similar among a number of users. For our project, we used the Spotify popularity score feature. The score is assigned to all the songs in our dataset in range between 0 and 100, with 100 being the most popular. The score reflects the overall popularity of a song's artist on the platform.
After sorting the output of the previous step by this popularity score, we finally get the recommendation playlist in Figure 6 that we can let our test users to evaluate.

\section{Evaluation Method and Results}
\begin{figure}
  \centering
  \includegraphics[width=0.9\linewidth]{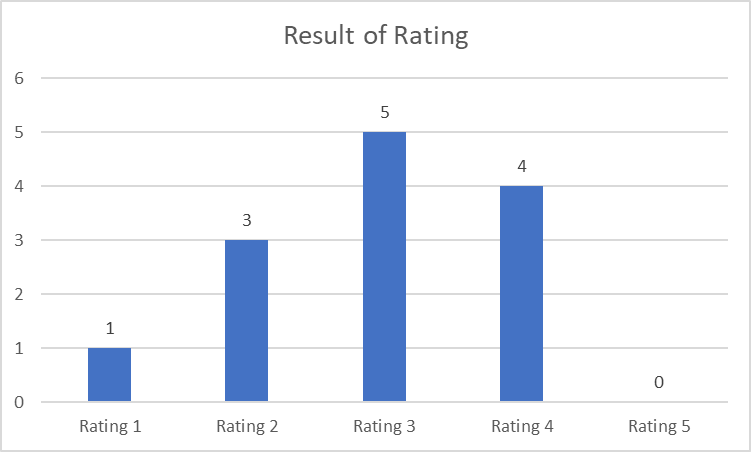}
  \caption{\textbf{Feedback results}: The graph shows the ratings distributions, with 1 indicating the least satisfaction with generated playlist, and 5 the most satisfaction.}
  \label{fig:sfig2}
\end{figure}
Since our project is related more to human psychology and preference, we do not want to use any statistical parameter to evaluate how well we are doing with this system. On the other hand, we decided to invite some of our friends and people who are willing to try and rate our recommendation algorithm results. \par
We gathered a dozen of people and asked them to provide us with three of their recent favorite songs. Since there are a lot of songs with the same name, we also gathered the artist name for the three songs to make sure that the song they pick is correctly entered into the system. We also gathered the data from ourselves in order to get more accurate evaluation results. \par
After we retrieved this data, we entered the songs in our system and created a 9 songs playlist for everyone, and let them evaluate the result. We asked them to rate our system’s performance on a scale from 1 to 5 which 1 is very poor, and 5 is very good. We also asked them to give us some comments including what part did the system do a good job, what part did the system do a bad job, and some other comments. Figure 5 is a bar chart of their ratings distribution. \par
We retrieved an average rating of 2.92 and we gathered a lot of feedback. The participants thought that we did a good job on rock and pop music. We also did a great job in identifying the tempo and loudness well. One of our testers liked quiet music and the playlist, we returned to them, is all quiet songs, which he really enjoyed listening to. On the other hand, we also got comments on the parts our algorithm performed badly on. Our system failed to identify electronic dance music or EDM. And also, some of the songs we recommend are too old to be appreciated by current users. 

\begin{figure*}
  \centering
  \includegraphics[width=0.4\linewidth]{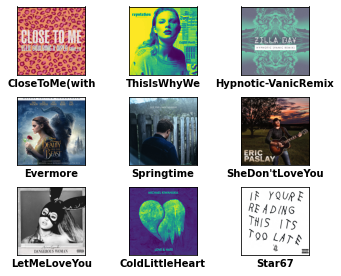}
  \caption{\textbf{Recommended playlist}: the result of our combined content and collaborative techniques is a playlist of the top K songs taken from the system's output.}
  \label{fig:sfig2}
\end{figure*}

\section{Conclusion and Future Work}

We judged an average rating of 2.92 as an alright job for a start because people judged the songs not only based on arousal and valence, which are the two factors we used to make recommendations for others. There are a lot more factors that could impact the judgment such as the mood they are in and their popularity. There are also songs that you might have no feeling when you first listen to them, we got no feelings about them. But as we listen to them more times, we are addicted to them. Whereas there are also songs when you first listen to them, you feel really addicted but as you listen to them more times, you get tired of them. So it is hard to retrieve a high rating on the song recommendation system. But based on the comments, we have set a couple of directions we could work on. \par
The first thing we wanted to improve on is to change some of the parameters. In the beginning, we limit the result of the output cannot be older than 8 years for the song we put into the system. Since we get the result that the song recommended is too old, we will try to decrease the year boundary to 5 or even less. We will also add the genre classification to our tracks. The reason we haven't done that Spotify does not assign genres to the songs, but to the artists which makes it harder and more subjective to math songs and genres. And also, Spotify limits the number of songs we can retrieve to build our database. However, we will look into more datasets and companies to retrieve a more comprehensive dataset with more classification methods to make the recommendation.

Another thing we want to improve on is the input. Currently, we used the song’s Spotify ID as input it will really mess us up when we have a playlist with a lot of songs in it. So we want to try changing the input as a playlist so it will be convenient for us and the users. And with more songs, we can better retrieve the users’ preferences and make better recommendations. 

The last thing we want to improve on is emotional analysis. We will look for research tools to collect users’ emotional feedback data such as MuPsych which has been integrated into Spotify. The reason we want to make this improvement is so we can better understand the emotional status and have a deeper understanding of why the user is addicted to this kind of song. As we stated earlier, it is hard for us to get a high rating when recommending songs because there are a lot of emotional factors that might influence the result. However, with the addition of emotional status research tools, we could understand the users’ emotional factors and we can make recommendations with higher accuracy to them.

\bibliographystyle{plain}
\bibliography{citations.bib}

\end{document}